\documentclass[aps,pra,amssymb,superscriptaddress,footinbib,twocolumn]{revtex4}
\usepackage{graphicx}
\usepackage{amsmath}
\usepackage{color}
\usepackage{ulem} %to strike out
\usepackage{marginnote}
\usepackage{tikz}

\begin{document}

\title{Generation of quantum steering and interferometric power \\ in the Dynamical Casimir Effect}

 \author{Carlos Sab\'in}
\email{carlos.sabin@nottingham.ac.uk}
\affiliation{School of Mathematical Sciences, The University of Nottingham,
  Nottingham NG7 2RD, United Kingdom}
\author{Gerardo Adesso}
\affiliation{School of Mathematical Sciences, The University of Nottingham,
  Nottingham NG7 2RD, United Kingdom}
  \date{\today}

\begin{abstract}

We analyse the role of the dynamical Casimir effect as a resource for quantum technologies, such as quantum cryptography and quantum metrology. In particular, we consider the generation of Einstein-Podolsky-Rosen steering and Gaussian interferometric power, two useful forms of asymmetric quantum correlations, in superconducting waveguides modulated by superconducting quantum interferometric devices. We show that, while a certain value of squeezing is required to overcome thermal noise and give rise to steering, any non-zero squeezing produces interferometric power which in fact increases with thermal noise.
\end{abstract}
\maketitle

\section{Introduction}

The first experimental observation of the Dynamical Casimir Effect (DCE)  \cite{moore} in a superconducting circuit architecture in 2011  \cite{casimirwilson} has triggered a renewed interest in this phenomenon. The DCE consists in the generation of photons out of the vacuum of a quantum field by means of the relativistic motion of boundary conditions. Letting alone its paramount foundational relevance as a paradigmatic prediction of relativistic quantum field theory, the fact that it can be realised by the modulation of a superconducting Quantum Interferometric Device (SQUID) interrupting a superconducting transmission line has paved the way for the analysis of the role of DCE radiation as a resource for quantum technologies \cite{benenti}.

In particular, it has been shown that the photon pairs generated by the DCE display quantum entanglement \cite{nonclassicaldce} and quantum discord \cite{mypra} under realistic experimental conditions. Moreover, these correlations can be swapped to superconducting qubits, enabling the generation of highly entangled qubit states \cite{felicetti}. It is thus of particular relevance to investigate in more detail how the DCE can be exploited to generate other useful forms of quantum correlations, and how robust they are to thermal noise in realistic settings.

In mixed states of bipartite systems, one can distinguish in fact different layers of quantum correlations, which in order of decreasing strength include: Bell nonlocality \cite{nonlocality}, Einstein-Podolsky-Rosen (EPR) steering \cite{wiseman}, quantum entanglement \cite{horodecki}, and discord-type correlations \cite{modietal}. In this paper we focus on two correlation measures, both having an asymmetric nature, and both playing resource roles for some important quantum technological tasks involving Gaussian states of continuous variable systems.

On one hand, we consider a computable measure of Gaussian quantum steering  $\cal{G}^{A \rightarrow B}$, which captures the EPR paradox, and quantifies to which extent Bob's mode can be steered by Alice's Gaussian measurements on her mode, in a two-mode entangled Gaussian state \cite{steering}. Operationally, the Gaussian quantum steering measure can be interpreted as the guaranteed key rate achievable in a one-way device-independent quantum key distribution protocol based on shared Gaussian states and reverse reconciliation \cite{walk}.

On the other hand, we consider a metrological figure of merit that captures a form of discord-type correlations, known as interferometric power \cite{discordtech3}. In the continuous variable setting, the Gaussian interferometric power ${\cal P}_G^A$ is given by the minimum quantum Fisher information for estimating a phase shift applied to Alice's probe mode in an optical interferometer, minimised over the (Gaussian) generators of the transformation encoding the phase \cite{interferometricpower,manab}. Operationally, the Gaussian interferometric power can be interpreted as the guaranteed precision achievable by a two-mode Gaussian state for unitary parameter estimation in an interferometric configuration.

Both quantities are therefore directly relevant for practical applications, and both admit simple closed formulae for two-mode Gaussian states, a fact which will be advantageous for our analysis. We remark that any separable or entangled Gaussian state can have a nonzero interferometric power \cite{interferometricpower}, while only a subset of entangled states can be steered by Gaussian measurements \cite{wiseman,steering}.

In the following, we compute both the Gaussian quantum steering and the Gaussian interferometric power generated in the experimental setup employed for the observation of the DCE, namely a superconducting waveguide interrupted by a SQUID \cite{casimirwilson}.
Using realistic experimental parameters, we show that these quantities exhibit quite different features. In the case of quantum steering, the value of the experimental driving amplitude and velocity must be higher than a critical value in order to overcome the initial level of thermal noise, a similar behaviour to the one predicted for entanglement and other quantum correlations \cite{nonclassicaldce, mypra}. Conversely, the interferometric power is non-zero for any experimental value of the amplitude and velocity, regardless of the level of thermal noise. Remarkably, it increases with the average number of thermal phonons. This ties in with the observation that the performance of quantum metrology, discrimination and reading protocols with continuous variable probes can indeed be enhanced by thermal noise \cite{interferometricpower,faberreading}.

The structure of the paper is the following. In section II we briefly recall the formalism of the Dynamical Casimir Effect in superconducting circuits, and in particular we compute the covariance matrix of the system. In section III we show the results for both gaussian quantum steering and gaussian interferometric power. We conclude in section IV with a summary of the results and a discussion of possible applications.

\section{Dynamical Casimir Effect with superconducting circuits}

We will consider the same experimental setup as in \cite{casimirwilson, nonclassicaldce}. The electromagnetic field confined by a superconducting waveguide is described by a quantum field associated to the flux operator $\Phi(x,t)$,
%It is related to the voltage operator by $\Phi(x,t) = \int^tdt'V(x,t')$, and to the gauge-invariant superconducting phase operator $\varphi = 2\pi\Phi/\Phi_0$, where $\Phi_0 = h/2e$ is the magnetic flux quantum.
which obeys a (1+1)-dimensional Klein-Gordon wave equation, $\partial_{xx}\Phi(x,t)-v^{-2}\partial_{tt}\Phi(x,t)=0$. The  field can thus be written as:
\begin{eqnarray}
\label{eq:field}
\Phi(x,t) &=& \sqrt{\frac{\hbar Z_0}{4\pi}}\int_{-\infty}^{\infty} \frac{d\omega}{\sqrt{|\omega|}}\times\\
&&\nonumber
\left[a(\omega) e^{-i(-k_\omega x +\omega t)} + b(\omega)e^{-i(k_\omega x +\omega t)}\right],
\end{eqnarray}
where $a(\omega)$ and $b(\omega)$ are the annihilation operators for photons with frequency $\omega$ propagating to the right (incoming) and left (outgoing), respectively. Here  $k_\omega = \omega/v$ is the wavenumber, $v$ is the speed of light in the waveguide, $Z_0$ is the characteristic impedance, and we have used the notation $a(-\omega)=a^\dag(\omega)$.

As shown in \cite{johansson:2009,johansson:2010}, for large enough SQUID plasma frequency, the SQUID is a passive element that provides the following boundary condition to the flux field:
\begin{eqnarray}
\Phi(0,t) + \left.L_{\rm eff}(t)\partial_x\Phi(x,t)\right|_{x=0} = 0,
\end{eqnarray}
that can be described by an effective length
\begin{equation}
L_{\rm eff}(t) = \left(\Phi_0/2\pi\right)^2/(E_J(t)L_0),
\end{equation}
where $L_0$ is the characteristic inductance per unit length of the waveguide and $E_J(t)=E_J[\Phi_{\rm ext}(t)]$ is the flux-dependent effective Josephson energy.  For sinusoidal modulation with driving frequency $\omega_d/2\pi$ and normalized amplitude $\epsilon$, $E_J(t) = E_J^0 [1 + \epsilon \sin \omega_d t]$, we obtain an effective length modulation amplitude $\delta\!L_{\rm eff} = \epsilon L^0_{\rm eff}$, where $L^0_{\rm eff} = L_{\rm eff}(0)$. If the effective velocity $v_{\rm eff}=\delta\!L_{\rm eff}\omega_d$ is a significant fraction of $v$, the emission of DCE photon pairs is sizeable.

The DCE can be analysed using scattering theory which describes how the time-dependent boundary condition mixes the otherwise independent incoming and outgoing modes \cite{lambrecht:1996}.
In the perturbative regime discussed analytically in \cite{johansson:2009,johansson:2010,nonclassicaldce}, the resulting output field is correlated to modes with frequencies $\omega_+, \omega_-$, such that $\omega_+ + \omega_-=\omega_d$, so we can write $\omega_\pm = \omega_d/2 \pm \delta\omega$, where $\delta\omega$ is the detuning.
Introducing the notation $a_\pm=a(\omega_\pm)$ and $b_\pm=b(\omega_\pm)$, the relation between the input and the output fields is the following:
\begin{eqnarray}
\label{eq:output-field-perturbation-simplified-notation}
b_\pm = -a_\pm -i\frac{\delta\!L_{\rm eff}}{v}\sqrt{\omega_+\omega_-}a^\dag_\mp,
\end{eqnarray}
where  $\delta\!L_{\rm eff}\sqrt{\omega_-\omega_+}/v$ is a small parameter. If we consider small detuning, then $\omega_-\simeq \omega_+\simeq\omega_d/2$ and
\begin{equation}
\frac{\delta\!L_{\rm eff}\sqrt{\omega_-\omega_+}}{v}\simeq\frac{\epsilon L_{\rm eff}(0)\omega_d}{2v}=\frac{v_{\rm{eff}}}{2v}.
\end{equation}
Denoting the small parameter as $f$, we can write:
\begin{eqnarray}
\label{eq:output-field-perturbation-simplified-notation2}
b_\pm = -a_\pm -i\,f\,a^{\dag}_\mp.
\end{eqnarray}

Let us consider now the covariance matrix of the system $V$. Using the same convention as in \cite{nonclassicaldce}, which assumes zero displacement without any loss of generality, we have $$V_{\alpha\beta} = \frac{1}{2}\left<R_\alpha R_\beta+R_\beta R_\alpha\right>,$$  where $R^{\rm T} = \left(q_-, p_-, q_+, p_+\right)$ is a vector with the quadratures as elements: $q_\pm = (b_\pm + b_\pm^\dag)/\sqrt{2}$ and $p_\pm = -i(b_\pm - b_\pm^\dag)/\sqrt{2}.$ Note that the quadratures of the outgoing modes can be written in terms of those of the ingoing modes, $q_{0\pm} = (a_\pm + a_\pm^\dag)/\sqrt{2}$ and $p_{0\pm} = -i(a_\pm - a_\pm^\dag)/\sqrt{2}$, by using
Eq.~(\ref{eq:output-field-perturbation-simplified-notation2}):
\begin{equation}
\label{eq:cuadratures}
q_\pm=-(q_{0\pm}+\,f\,p_{0\mp})\,,\quad
p_\pm=-(p_{0\pm}+\,f\,q_{0\mp}).
\end{equation}
We assume that the ingoing modes are in a weakly thermal, quasi-vacuum state characterised by a small fraction of thermal photons $n^{\rm{th}}_+$, $n^{\rm{th}}_-$ as is the case for typical $\rm{GHz}$ frequencies and $\rm{mK}$ temperatures  in a superconducting scenario. Then the ingoing covariance matrix is
\begin{equation}
\label{eq:covariancematrixin}
V_0=\frac{1}{2} \begin{pmatrix} 1+2\,n^{\rm{th}}_-&0&0&0\\0&1+2\,n^{\rm{th}}_-&0&0\\ 0&0&1+2\,n^{\rm{th}}_+&0\\0&0&0&1+2\,n^{\rm{th}}_+ \end{pmatrix}\!.
\end{equation}
Note that since we are considering small detuning, $\omega_+\simeq\omega_-\simeq\omega_d/2$, it follows that $n^{\rm{th}}_+\simeq n^{\rm{th}}_-\simeq n^{\rm{th}}$, that is, the states can be considered approximately symmetric under a swap of the two modes. Using Eqs.~(\ref{eq:cuadratures}) and (\ref{eq:covariancematrixin}), we obtain the covariance matrix of the outgoing modes
\begin{eqnarray}\label{eq:covariancematrixout}
V &=&\frac{1}{2} \begin{pmatrix} A & C\\ C^T & B\end{pmatrix}, \\
A&=&1+2\,n^{\rm{th}}_- +f^2 (1+2\,n^{\rm{th}}_+)\openone,\nonumber\\
B&=&1+2\,n^{\rm{th}}_+ +f^2 (1+2\,n^{\rm{th}}_-)\openone, \nonumber\\
C&=&2f (1+ n^{\rm{th}}_+ + n^{\rm{th}}_-)\sigma_x.\nonumber
\end{eqnarray}

This is a two-mode squeezed thermal state characterised by the squeezing parameter $2f\,$ and its standard form is obtained just by replacing the Pauli matrix $\sigma_x$ with $\sigma_z$ in $C$.

\section{Quantum steering and interferometric power}

We are now ready to characterise the aforementioned two different forms of quantum correlations for the state described by the covariance matrix in Eqs.~(\ref{eq:covariancematrixout}). We start by computing the Gaussian quantum steering, which for Gaussian bipartite states takes the form \cite{steering}:
\begin{equation}\label{eq:Gaussiansteering}
\mathcal{G}^{A\rightarrow B} (V)=\operatorname{max} \left\{0,\frac{1}{2}\operatorname{log}\frac{\operatorname{det}A}{\operatorname{det}V} \right\}.
\end{equation}
Using Eqs.~(\ref{eq:Gaussiansteering}) and (\ref{eq:covariancematrixout}) we can evaluate the Gaussian quantum steering in the perturbative regime. We find:
\begin{equation}\label{eq:steeringdce}
\mathcal{G}^{A\rightarrow B} (V)=\operatorname{max} \{0, 3\,f^2-2\,n^{\rm{th}} \}.
\end{equation}
Therefore, the DCE radiation exhibits non-zero Gaussian steering as long as
\begin{equation}\label{eq:condition}
f>\sqrt{\frac{2\, n^{\rm{th}}}{3}}.
\end{equation}
Writing it explicitly in terms of the driving amplitude $\epsilon$, we find that the onset of the Gaussian quantum steering occurs at:
\begin{equation}\label{eq:onset}
\epsilon_0= \frac{2\,\sqrt{2}\,v\,f}{\sqrt{3}\, L_{\rm eff}(0)\omega_d}\,\sqrt{n^{\rm{th}}}.
\end{equation}

Let us analyse now the behaviour of the Gaussian interferometric power, whose expression is given by  \cite{interferometricpower}:
\begin{equation}\label{eq:ip}
\mathcal{P}^A_G (V)=\frac{X+\sqrt{X^2+YZ}}{2Y},
\end{equation}
where
\begin{eqnarray}
X=(I_1+I_3)(1+I_2+I_3-I_4)-I_4^2\nonumber\\
Y=(I_4-1)(1+I_1+I_2+2\,I_3+I_4)\\
Z=(I_1+I_4)(I_1I_2-I_4)+I_3(2I_1+I_3)(1+I_2),\nonumber
\end{eqnarray}
with
$$I_1=\det A,\, I_2=\det B,\, I_3=\det C,\,I_4=\det V.$$ Using Eqs.~(\ref{eq:ip}) and (\ref{eq:covariancematrixout}) we find that in the perturbative regime the interferometric power of the DCE state is:
\begin{equation}\label{eq:ipdce}
\mathcal{P}^A_G (V)=f^2(1+2\,n^{\rm{th}}).
\end{equation}
Therefore we note that, regardless the degree of thermal noise, the interferometric power is generated for any non-zero value of the squeezing parameter, and it further increases with the number of thermal photons.
\begin{figure}[h!]
\includegraphics[width=\columnwidth]{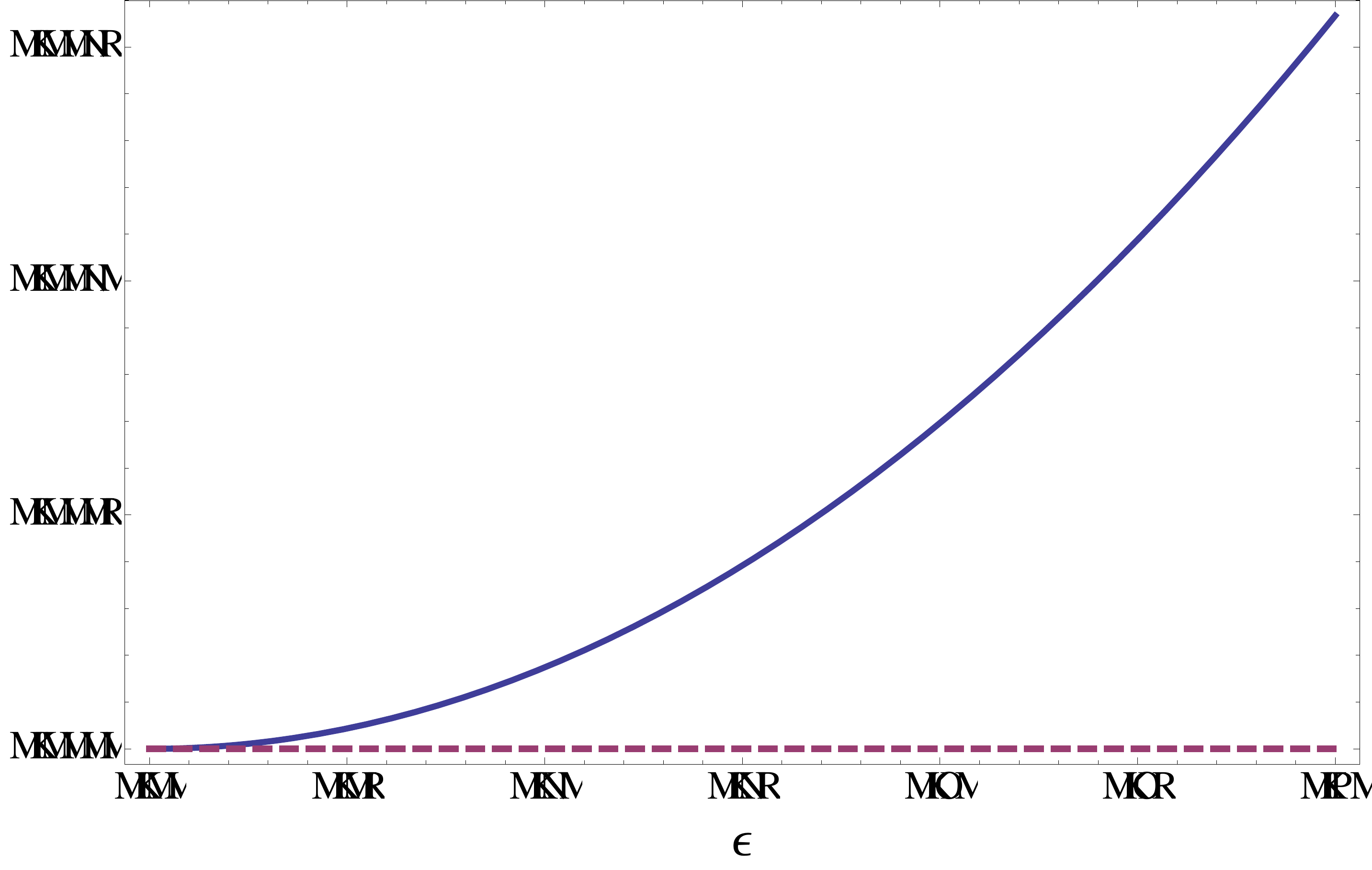}
\caption{\label{fig:fig1} (Color online) Gaussian interferometric power -blue, solid- and Gaussian quantum steering -purple, dashed- as a function of the normalised driving ampitude $\epsilon$. We consider experimental parameters $v=1.2\cdot 10^8 \rm{m/s}$, $\omega_d=2\pi\cdot 10\rm{GHz}$, $L_{\rm eff}(0)=0.5\rm{mm}$ and $T=50 \rm mK$. Thus the small parameter $f<0.05$ is well within the perturbative regime, as well as the average numbers of thermal photons $n^{\rm th}\simeq 8\cdot 10^{-3}$. Quantum steering is 0 in this regime while interferometric power increases quickly with the driving amplitude.}
\end{figure}
\begin{figure}[h!]
\includegraphics[width=\columnwidth]{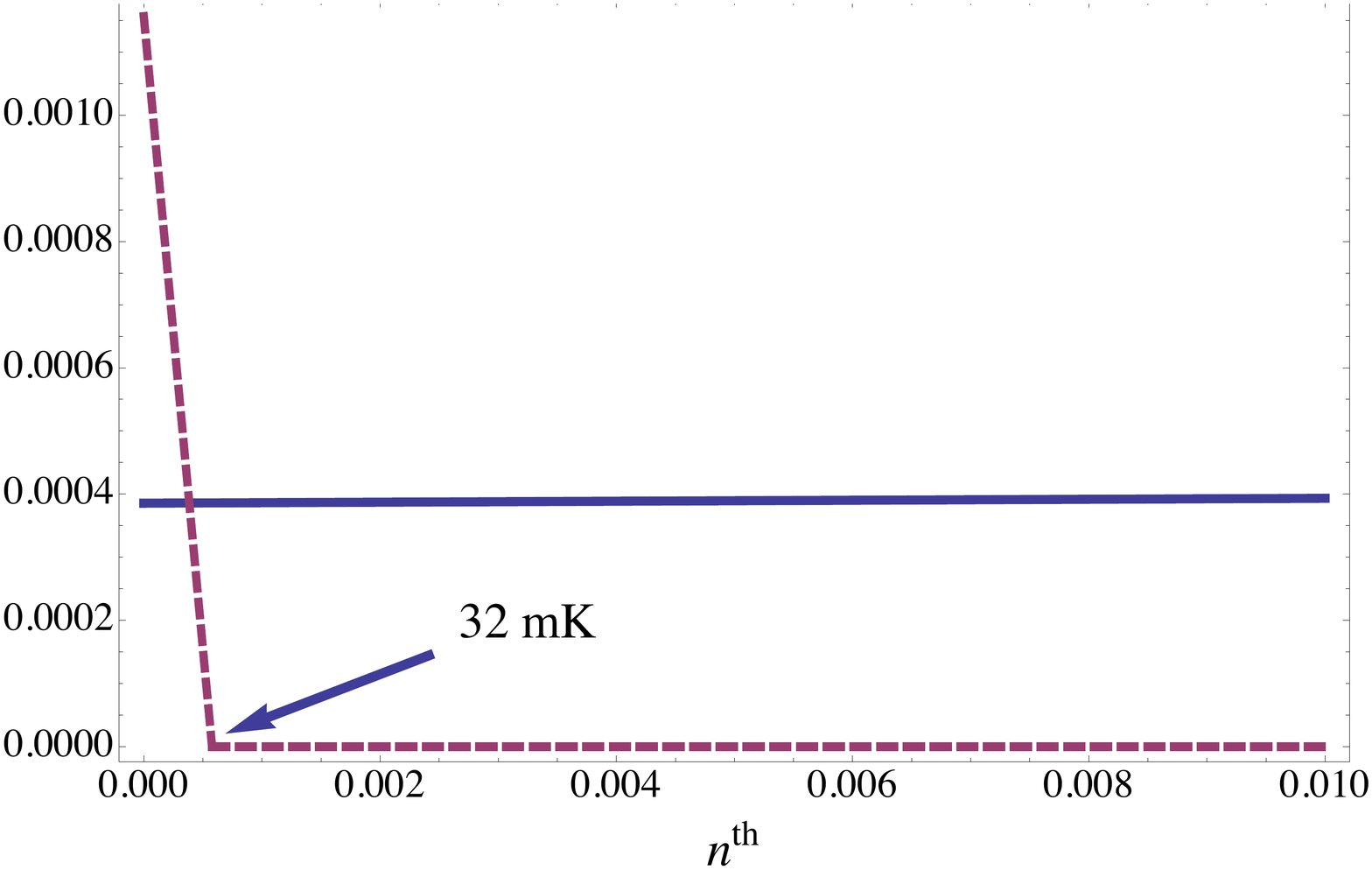}
\caption{\label{fig:fig2}
(Color online)Gaussian interferometric power -blue, solid- and Gaussian quantum steering -purple, dashed- as a function of the average number of thermal photons $n^{\rm th}$. We consider experimental parameters $v=1.2\cdot 10^8 \rm{m/s}$, $\omega_d=2\pi\cdot 10\rm{GHz}$, $L_{\rm eff}(0)=0.5\rm{mm}$ and $\epsilon=0.15$. Thus the small parameter $f\simeq0.02$ is well within the perturbative regime. Quantum steering quickly decreases with temperature and vanishes at $T\simeq32\rm mK$. In contrast, the interferometric power is always nonzero and increases with temperature.}
\end{figure}

In Figs.~\ref{fig:fig1} and \ref{fig:fig2} we analyse the behaviour of quantum steering and interferometric power with respect to the driving amplitude and average number of thermal photons in a realistic experimental regime.
We observe that for a realistic temperature of $T=50\,\operatorname{mK}$ \cite{casimirwilson} quantum steering is 0 for any sensible value of $\epsilon$. Indeed, quantum steering is very fragile to thermal noise and is only different from zero below $32\,\operatorname{mK}$ for the value of $\epsilon$ achieved in \cite{casimirwilson}, in contrast to entanglement and entropic discord which display critical thresholds of $60$ and $67\,\operatorname{mK}$ respectively for the same value of $\epsilon$. On the other hand, the interferometric power achieves finite values that increase quickly with the driving amplitude and are almost insensitive to thermal noise in this regime.

\begin{figure}[t!]
\includegraphics[width=\columnwidth]{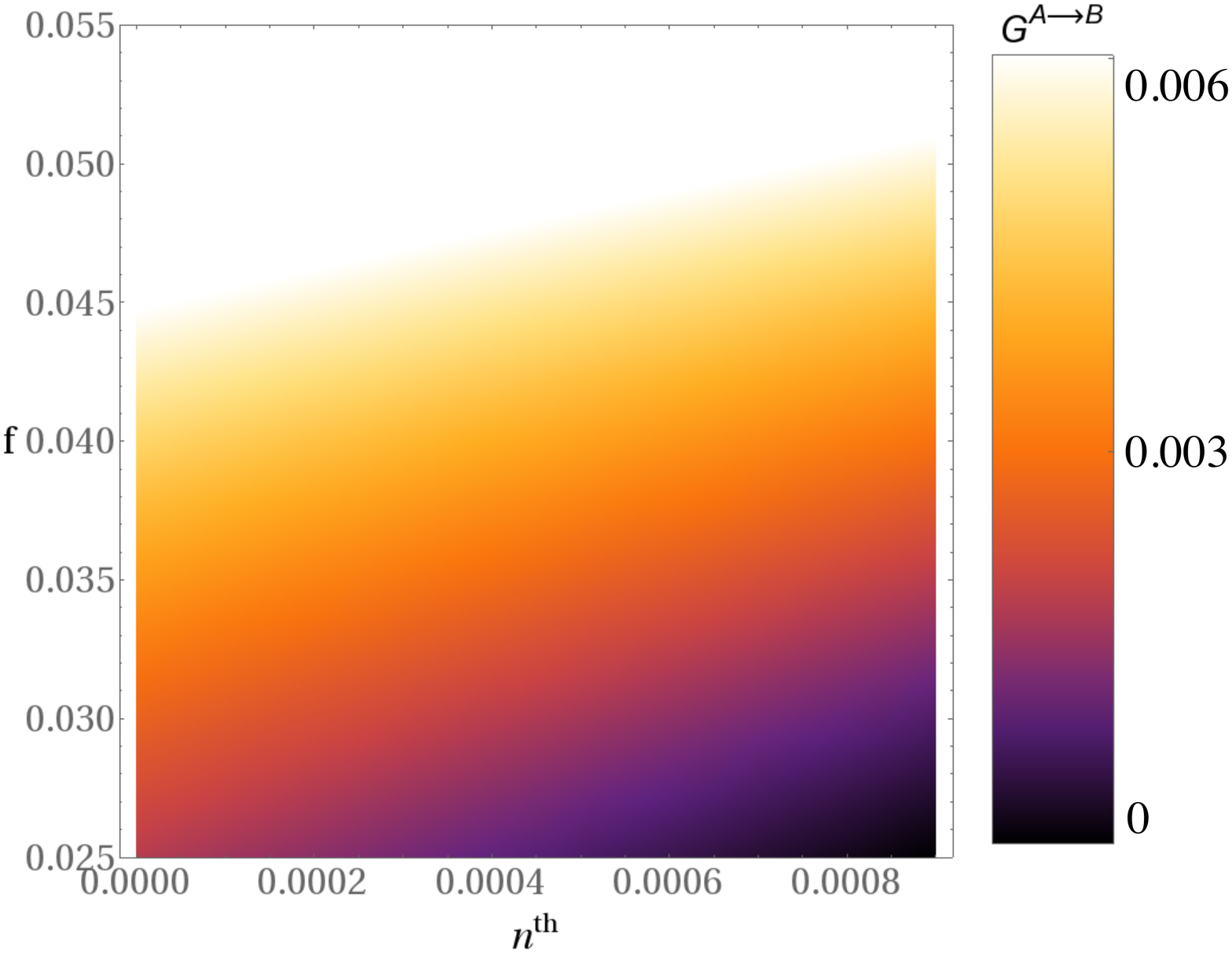}
\caption{\label{fig:fig3}
(Color online) Gaussian quantum steering as a function of the average number of thermal photons $n^{\rm th}$ and the squeezing parameter $f$. The number of photons considered corresponds to temperatures ranging from 0 to $35\,\operatorname{mK}$. Steering achieves non-zero values in this regime of temperatures.}
\end{figure}

Finally, in order to understand better the interplay between squeezing and temperature in the generation of quantum steering, we plot the measure ${\cal G}^{A \rightarrow B}$  in Fig.~\ref{fig:fig3} in the parameter regime where it is non-zero. For sufficiently high values of the squeezing, quantum steering survives for temperatures as big as $35 \,\operatorname{mK}$.

\section{Conclusions}
In conclusion, we have investigated the experimental scenario of the DCE demonstration in a superconducting waveguide terminated by a SQUID, focusing our attention on the nature of the correlations in the generated radiation. 

We studied in particular the ability of the DCE for the generation of EPR quantum steering \cite{steering}, a form of quantum correlations stronger than entanglement and essential for one-way device-independent quantum cryptography \cite{wiseman,walk}, and of interferometric power \cite{interferometricpower}, a form of quantum correlations weaker than entanglement, which captures the usefulness of a state to act as a probe for quantum metrology in a worst-case scenario \cite{discordtech3}. 

We found that both correlations can be generated by the DCE in realistic experimental conditions, although they exhibit quite different behaviours. On one hand, steering is very fragile and disappears for moderate levels of thermal noise, even though the state of the radiation may remain entangled. On the other hand, interferometric power is always nonvanishing and is even enhanced by thermal noise. This shows that the DCE can be regarded as an effective and practical resource to generate useful correlations for entanglement-based \cite{felicetti} and non-entanglement-based quantum technologies in the continuous variable setting, including in particular quantum estimation and communication. 
An experimental verification of black-box phase estimation \cite{interferometricpower} in a superconducting architecture \cite{casimirwilson} or in a Bose-Einstein condensate \cite{casimirwestbrook}, based exclusively on correlated probe states of the (radiation or phononic) field generated by the DCE, and exploiting thermal enhancements, would be an intriguing subject for a future work. Finally, let us highlight that these results would be valid as well to any process that generates multimode squeezed states in the presence of thermal noise.

G. A. acknowledges financial support from the Foundational Questions Institute  and the University of Nottingham International Collaboration Fund.

\end{document}